\documentclass[11pt]{article}
\usepackage{graphicx}
\usepackage{epsfig}
\usepackage{amsfonts}

\usepackage{amsmath}

\newcommand{\beq}{\begin{equation}}
\newcommand{\eeq}{\end{equation}}
\newcommand{\beqs}{\begin{eqnarray}}
\newcommand{\eeqs}{\end{eqnarray}}

\begin{document}
\title{AdS$_5$ rotating non-Abelian black holes}
\author{{\large Yves Brihaye,}$^{\ddagger}$
{\large Eugen Radu}$^{\dagger}$
and {\large D. H. Tchrakian}$^{\dagger \star}$ \\ \\
$^{\ddagger}${\small Physique-Math\'ematique, Universite de
Mons-Hainaut, Mons, Belgium}\\
$^{\dagger}${\small Department of
Mathematical Physics, National University of Ireland Maynooth,} \\
{\small Maynooth, Ireland} \\
$^{\star}${\small School of Theoretical Physics -- DIAS, 10 Burlington
Road, Dublin 4, Ireland }}

\date{}
\newcommand{\dd}{\mbox{d}}
\newcommand{\tr}{\mbox{tr}}
\newcommand{\la}{\lambda}
\newcommand{\ka}{\kappa}
\newcommand{\f}{\phi}
\newcommand{\al}{\alpha}
\newcommand{\ga}{\gamma}
\newcommand{\de}{\delta}
\newcommand{\si}{\sigma}
\newcommand{\bomega}{\mbox{\boldmath $\omega$}}
\newcommand{\bsi}{\mbox{\boldmath $\sigma$}}
\newcommand{\bchi}{\mbox{\boldmath $\chi$}}
\newcommand{\bal}{\mbox{\boldmath $\alpha$}}
\newcommand{\bpsi}{\mbox{\boldmath $\psi$}}
\newcommand{\brho}{\mbox{\boldmath $\varrho$}}
\newcommand{\beps}{\mbox{\boldmath $\varepsilon$}}
\newcommand{\bxi}{\mbox{\boldmath $\xi$}}
\newcommand{\bbeta}{\mbox{\boldmath $\beta$}}
\newcommand{\ee}{\end{equation}}
\newcommand{\eea}{\end{eqnarray}}
\newcommand{\be}{\begin{equation}}
\newcommand{\bea}{\begin{eqnarray}}
\newcommand{\ii}{\mbox{i}}
\newcommand{\e}{\mbox{e}}
\newcommand{\pa}{\partial}
\newcommand{\Om}{\Omega}
\newcommand{\vep}{\varepsilon}
\newcommand{\bfph}{{\bf \phi}}
\newcommand{\lm}{\lambda}
\def\theequation{\arabic{equation}}
\renewcommand{\thefootnote}{\fnsymbol{footnote}}
\newcommand{\re}[1]{(\ref{#1})}
\newcommand{\R}{{\rm I \hspace{-0.52ex} R}}
\newcommand{\N}{{\sf N\hspace*{-1.0ex}\rule{0.15ex}%
{1.3ex}\hspace*{1.0ex}}}
\newcommand{\Q}{{\sf Q\hspace*{-1.1ex}\rule{0.15ex}%
{1.5ex}\hspace*{1.1ex}}}
\newcommand{\C}{{\sf C\hspace*{-0.9ex}\rule{0.15ex}%
{1.3ex}\hspace*{0.9ex}}}
\newcommand{\eins}{1\hspace{-0.56ex}{\rm I}}
\renewcommand{\thefootnote}{\arabic{footnote}}

\maketitle

%\ \ \ PACS Numbers: 04.50.+h, 11.10.Kk, 11.15.Kc

\bigskip

\begin{abstract}
We present arguments for the existence of charged,
rotating black holes
with equal magnitude angular momenta in $d=5$ Einstein-Yang-Mills theory
with negative cosmological constant.
These solutions posses
a regular horizon of spherical topology and
approach asymptotically the Anti-de Sitter spacetime background.
The black hole solutions have also an electric
charge and a nonvanishing magnetic flux through the sphere at infinity.
Different from the static case, no regular solution with a
nonvanishing angular momenta is found
for a vanishing event horizon radius.
\end{abstract}
\medskip

%%%%%%%%%%%%%%%%%%%%%%%%%%%%%%%%%%%%%%%%%%%%%%%%%%%%%%%%%%%%%%%%%%
\section{Introduction}
%%%%%%%%%%%%%%%%%%%%%%%%%%%%%%%%%%%%%%%%%%%%%%%%%%%%%%%%%%%%%%%%%
The conjectured equivalence of string theory on anti-de Sitter (AdS)
spaces and certain superconformal gauge theories living on the
boundary of AdS \cite{Witten:1998qj,Maldacena:1997re} has lead
recently to an increasing interest in asymptotically anti-de Sitter
(AAdS) black holes. This type of solutions are of special interest
since they offer the possibility of studying the nonperturbative
structure of some conformal field theories (CFTs).

It is therefore desirable to widen the existing
 AAdS classes of solutions as much as possible,
the case of five dimensional solutions being of particular interest,
given the conjectured equivalence between $\mathcal{N}=4$, $d=4$ SU(N) YM theory
and supergravity solutions in AdS$_5$.
 Rotating black holes with AdS asymptotics in $d=5$
have been studied by various authors,
starting with Hawking ${\it et. al.}$ \cite{Hawking:1998kw}
who found the higher dimensional counterparts of the Kerr-AdS$_4$ solution.
Charged rotating black holes  solutions in $d=5$ gauged supergravity
have been constructed in \cite{Chong:2005hr}.
 Apart from Abelian fields with a Chern-Simons term, these configurations
usually contain scalar fields with a nontrivial scalar potential.
Rotating black hole solutions in pure Einstein-Maxwell (EM) theory
with negative cosmological constant have been constructed
numerically in recent work  \cite{Kunz:2007jq}.

At the same time, one should remark that gauged supergravity theories playing
an important role in AdS/CFT, generically contain non-Abelian matter fields in
the bulk, although in the literature mainly Abelian truncations are considered,
to date.
The lack of attention given to AAdS Einstein-Yang-Mills (EYM) solutions is presumably due
to the notorious absence of closed form solutions in this case.
However, one can analyse their
properties by using a combination of analytical and numerical
methods, which is enough for most purposes.
Thus, the examination of AdS$_5$ gravitating
non-Abelian solutions with $\Lambda<0$ is a pertinent task.

Practically, all that is known in the subject of
$d=5$ AAdS non-Abelian solutions are the EYM-SU(2) spherically symmetric
configurations discussed in \cite{Okuyama:2002mh}
and the static solutions in \cite{Radu:2006va}
of the $N=4^+$ version of the Romans' gauged supergravity model \cite{Romans:1985ps}.
These  solutions share a number of properties with the better
known $d=4$ EYM AAdS configurations discussed in
\cite{Winstanley:1998sn},\cite{Bjoraker:2000qd}.
In both cases,
regular and black hole solutions exist for compact intervals
of the parameter
that specifies the initial conditions at the origin or at the event horizon.
The gauge field
approaches asymptotically a configuration  which is not
a pure gauge, resulting in a nonvanishing magnetic
flux through the sphere at infinity.

However, different from the four dimensional case,
the mass of an AdS$_5$ configuration as defined in the usual way
presents a logarithmic divergence.
In the recent work \cite{BRT}, a counterterm based method has been proposed to
regularise  the action and mass-energy of the non-Abelian AdS$_5$ solutions.
In this approach,
the logarithmic divergence of the action results in a trace anomaly in the dual CFT.

The main purpose of this paper
is to   present  numerical
arguments for the existence of rotating
 AAdS$_5$ non-Abelian black hole solutions.
Instead
of specializing to a particular supergravity model, we shall consider the
simpler case of a EYM-SU(2) theory with negative cosmological constant.
Although it seems that this theory is not a consistent truncation
of any $d=5$ supersymmetric model, it enters the gauged supergravities as the basic
building block and  one can expect the basic features of its solutions to be
generic.
Also, we shall restrict here to the case of rotating solutions with
equal magnitude angular momenta  and a spherical topology of the
event horizon,
which allows us to deal with ordinary differential equations (ODEs).

These solutions share a number of common features with the $d=4$
counterparts discussed in \cite{Mann:2006jc}, \cite{Radu:2002rv}.
In both cases, one finds rotating solutions starting with any static
configuration.
The rotating black holes have a nonzero electric charge and a nonvanishing flux
through the sphere at infinity.
However, different from the AdS$_4$ case \cite{Radu:2002rv}, here  no
rotating soliton solutions are found in the limit of zero event horizon radius.

The paper is structured as follows:  in Section 2 we present the general
framework and analyse the field equations and boundary conditions.
The black hole properties are discussed in Section 3.
We present the numerical results
in Section 4.
We conclude with Section 5 where the results are compiled.

%%%%%%%%%%%%%%%%%%%%%%%%%%%%%%%%%%%%%%%%%%%%%%%%%%%%%%%%%%%%%%%%%%
\section{The model}
%%%%%%%%%%%%%%%%%%%%%%%%%%%%%%%%%%%%%%%%%%%%%%%%%%%%%%%%%%%%%%%%%%
%%%%%%%%%%%%%%%%%%%%%%%%%%%%%%%%%%%%%%%%%%%%%%%%%%%%%%%%%%%%%%%%%%
\subsection{The action principle and field equations}
%%%%%%%%%%%%%%%%%%%%%%%%%%%%%%%%%%%%%%%%%%%%%%%%%%%%%%%%%%%%%%%%%%
We consider the five dimensional SU(2) Einstein-Yang-Mills (EYM)  action
with negative cosmological constant $\Lambda=-6/\ell^2$
\begin{eqnarray}
\label{action5}
I=\int_{\mathcal{M}} d^{5}x\sqrt{-g }\Big(\frac{1 }{16\pi G}(R-2\Lambda)
-\frac{1}{2e^2}{\rm Tr}\{F_{\mu \nu }F^{\mu \nu} \}\Big)
-\frac{1}{8 \pi G}\int_{\partial\mathcal{M}} d^4 x\sqrt{-h}K,
\end{eqnarray}
Here $G$ is the gravitational constant,
$R$ is the Ricci scalar associated with the
spacetime metric $g_{\mu \nu}$.
$F_{\mu \nu}=\frac{1}{2} \tau^aF_{\mu \nu}^{(a)}$ is the gauge field strength
tensor defined as
$F_{\mu \nu} =
\partial_\mu A_\nu -\partial_\nu A_\mu - i[A_\mu , A_\nu  ],
$
with a gauge potential
$A_{\mu} = \frac{1}{2} \tau^a A_\mu^{(a)},$
$\tau^a$ being the Pauli matrices and $e$ the gauge coupling
constant.
 $K$ is the trace
of the extrinsic curvature for the boundary $\partial\mathcal{M}$ and $h$ is
the induced  metric of the boundary.

Variation of the action (\ref{action5})
 with respect to  $g^{\mu \nu}$ and $A_\mu$ leads to the field equations
\begin{eqnarray}
\label{einstein-eqs}
R_{\mu \nu}-\frac{1}{2}g_{\mu \nu}R   = 8\pi G ~T_{\mu \nu},
~~
\nabla_{\mu}F^{\mu \nu}-i[A_{\mu},F^{\mu \nu}]=0,
\end{eqnarray}
where the YM stress-energy tensor is
\begin{eqnarray}
\label{tik} T_{\mu \nu} = \frac{2}{e^2}{\rm tr}\{
 F_{\mu \rho} F_{\nu \lambda} g^{\rho \lambda}
   -\frac{1}{4} g_{\mu \nu} F_{\rho \lambda} F^{\rho \lambda}\}.
\end{eqnarray}

%%%%%%%%%%%%%%%%%%%%%%%%%%%%%%%%%%%%%%%%%%%%%%%%%%%%%%%%%%%%%%%%%%
\subsection{The ansatz}
%%%%%%%%%%%%%%%%%%%%%%%%%%%%%%%%%%%%%%%%%%%%%%%%%%%%%%%%%%%%%%%%%%

While the
general EYM-AdS rotating black holes would possess two independent
angular momenta and a more general  topology
of the event horizon,
we  restrict here to configurations with
equal magnitude angular momenta and a spherical horizon topology.
The suitable metric ansatz reads  \cite{Kunz:2005nm}
\begin{eqnarray}
\label{metric}
 ds^2=\frac{dr^2}{f(r)}+g(r)d \theta^2+h^2(r)
(\sin^2 \theta(d \varphi-w(r)dt)^2
+\cos^2 \theta(d \psi-w(r)dt)^2 )
\\
\nonumber
-(h^2(r)-g(r))\sin^2 \theta \cos^2 \theta (d \varphi-d \psi)^2-f(r)\sigma^2(r) dt^2~,
\end{eqnarray}
where $\theta  \in [0,\pi/2]$, $(\varphi,\psi) \in [0,2\pi]$, $r$ and $t$ being the
radial and time coordinates.
This line element presents five Killing vectors
\begin{eqnarray}
\nonumber
&&K_1=\frac{1}{2}\sin (\psi-\varphi)\partial_\theta
-\frac{1}{2}\cos(\psi-\varphi)\cot \theta\partial_\varphi
-\frac{1}{2}\cos (\psi-\varphi)\tan \theta \partial_\psi,
\\
\label{KV}
&&K_2=\frac{1}{2}\cos (\psi-\varphi)\partial_\theta
+\frac{1}{2}\sin(\psi-\varphi)\cot \theta\partial_\varphi
+\frac{1}{2}\sin (\psi-\varphi)\tan \theta \partial_\psi,
\\
\nonumber
&&K_3=-\frac{1}{2}\partial_\varphi+\frac{1}{2}\partial_\psi,
~~K_4=\frac{1}{2}\partial_\varphi+\frac{1}{2}\partial_\psi,
~~K_5=\partial_t.
\end{eqnarray}
The computation of the appropriate  SU(2)  connection compatible
with the symmetries of the metric ansatz (\ref{metric})
can be done by applying the standard rules for calculating
the gauge potentials for any spacetime group
\cite{Forgacs:1980zs,Bergmann}.
According to Forgacs and Manton,
a gauge field admit a spacetime symmetry if the spacetime
transformation of the potential can be compensated by a gauge
transformation \cite{Forgacs:1980zs},
%\begin{eqnarray} \label{Psi}
$ {\mathcal{L}}_{K_i} A_{\mu}=D_{\mu}U_{i},$
%\end{eqnarray}
 where ${\mathcal{L}}$ stands for the Lie derivative.
The expression we find in this way for the gauge field ansatz is
\begin{eqnarray}
\label{ansatz-gauge}
A_r=0,~
A_\theta=\left( 2W(r),0,0\right )~,~
A_\varphi=\left(0,- W(r)\sin 2 \theta,  H(r) -\cos 2\theta (H(r)+1) \right)~,
\\
\nonumber
A_\psi=\left (0,W(r)\sin 2 \theta, H(r) +\cos 2\theta (H(r)+1) \right),~~
A_t=\left(0,0,V(r)\right)~,
\end{eqnarray}
the only nonvanishing components of the compensating potentials  $U_i$   being
\begin{eqnarray}
\label{U}
U_1=\frac{1}{2}\frac{\cos(\psi-\varphi)}{\sin 2 \theta}\tau_3,~~
U_2=\frac{1}{2}\frac{\sin(\psi-\varphi)}{\sin 2 \theta}\tau_3.
\end{eqnarray}
The general ansatz (\ref{metric}), (\ref{ansatz-gauge})
can be proven to be consistent, and,
as a result, the EYM equations reduce to a set of
seven ODEs (in the numerics, we fix the metric gauge
by taking $h(r)=r$).
The solutions have a spherically symmetric limit with
\begin{eqnarray}
\label{sph-limit}
g(r)=r^2,~h(r)=r,~ w(r)=0,~~W(r)=\frac{1}{2}({\tilde w}(r)+1),
~~~~H(r)=\frac{1}{2}({\tilde w}(r)-1),~~V(r)=0,
\end{eqnarray}
whose basic properties were discussed in \cite{Okuyama:2002mh}.
 The vacuum rotating black holes in \cite{Hawking:1998kw} with two equal angular
momenta
are recovered for a vanishing gauge field, $H=-1,~W=0$ (or, equivalently, $H=0,~W=1$),
 $V=0$
  and
\begin{eqnarray}
\label{vacuum}
f(r)=1
+\frac{r^2}{\ell^2}
-\frac{2{\hat M} }{r^{2}}(1-\frac{a^2}{\ell^2})
+\frac{2{\hat M}{\hat a}^2}{r^{4}},~~
h^2(r)=r^2\left(1+\frac{2{\hat M}{\hat a}^2}{r^{4}}\right),
\\
\nonumber
w(r)=\frac{2{\hat M}{\hat a}}{r h^2(r)},~~
g(r)=r^2,~~ b(r)=\frac{r^2 f(r)}{h^2(r)},
\end{eqnarray}
where ${\hat M}$ and ${\hat a}$ are two constants related to the solution's mass
and angular momenta.
The Einstein-Maxwell solutions in \cite{Kunz:2007jq}  are recovered for
the metric ansatz (\ref{metric}) written in
an isotropic coordinate system
and an U(1) subgroup of (\ref{ansatz-gauge}),
obtained for $W(r)\equiv 0$.

%%%%%%%%%%%%%%%%%%%%%%%%%%%%%%%%%%%%%%%%%%%%
\section{Black Hole Properties}
%%%%%%%%%%%%%%%%%%%%%%%%%%%%%%%%%%%%%%%%%%%%
%%%%%%%%%%%%%%%%%%%%%%%%%%%%%%%%%%%%%%%%%%%%%%%%%%%%%%%%%%%%%%%%
\subsection{Asymptotic expansion and boundary conditions }
%%%%%%%%%%%%%%%%%%%%%%%%%%%%%%%%%%%%%%%%%%%%%%%%%%%%%%%%%%%%%%%%
Similar to the vacuum case (\ref{vacuum}),
the horizon of these rotating black holes is a
squashed $S^3$ sphere and
resides at a constant value of the radial coordinate $r=r_h$,
being characterized by $f(r_h)=0$.
At the horizon, the solutions satisfy the boundary conditions
\begin{eqnarray}
\label{eh}
&& f|_{r=r_h }=0,~~g|_{r=r_h }=g_h,
~~\sigma|_{r=r_h }=\sigma_h,~~ w|_{r=r_h }= \Omega_H,~
\\
\nonumber
&&H|_{r=r_h }=H_h,~~W|_{r=r_h }=W_h,~~V|_{r=r_h }=-2\Omega_H H_h,
\end{eqnarray}
where $g_h,~\sigma_h,\Omega_H$, $H_h$ and $W_h$ are free parameters
(with $(g_h,~\sigma_h)>0$).

We find also the following asymptotic expansion as $r \to \infty$
\begin{eqnarray}
\label{a1} \nonumber
&&f(r)=1+\frac{r^2}{\ell^2}+\frac{f_2}{r^2}-\frac{128 \pi G }{e^2}
{(W_0-1)^2W_0^2}\frac{\log (r/\ell)}{r^2}+\dots, ~~
\sigma(r)=1+\frac{s_4}{r^4}+\dots,
\\
&&g(r)=r^2-\frac{s_4}{r^2}+\dots ,
~~
w(r)=\frac{\hat{J}}{r^4}+\dots,
\\
\nonumber
&&H(r)=W_0-1+\frac{H_2}{r^2}-2\ell^2 W_0(W_0-1)(2W_0-1)\frac{\log (r/\ell)}{r^2}+\dots,
\\
\nonumber
&&W(r)=W_0+\frac{W_2}{r^2}-2\ell^2 W_0(W_0-1)(2W_0-1)\frac{\log (r/\ell)}{r^2}+\dots,
~~V(r)=\frac{q}{r^2}+\dots,
\end{eqnarray}
where $f_2,s_4, \hat{J}$, $W_0,~W_2,~H_2$ and  $q$ are real constants.
Note that these asymptotics preserve the full AdS symmetry group.

One can see that, similar to the static case, the $g_{tt}$ component
of the metric has a term proportional with $(1-W_0)^2W_0^2(\log
r)/r^2$,
 which
leads to a divergent value of the mass-energy as defined in the
usual way, unless $W_0=0$ or $W_0=1$. However, we could not find
rotating non-Abelian solutions with these values of $W_0$. This
agrees with the physical intuition based on a heuristic Derick-type
scaling argument, although a rigorous proof exists for the
spherically symmetric limit only \cite{Okuyama:2002mh}.

%%%%%%%%%%%%%%%%%%%%%%%%%%%%%%%%%%%%%%%%%%%%%%%%%%%%%%%%%%%%%%%%
\subsection{Global charges }
%%%%%%%%%%%%%%%%%%%%%%%%%%%%%%%%%%%%%%%%%%%%%%%%%%%%%%%%%%%%%%%%
%%%%%%%%%%%%%%%%%%%%%%%%%%%%%%%%%%%%%%%%%%%%%%%%%%%%%%%%%%%%%%%%
\subsubsection{The mass and angular momenta}
%%%%%%%%%%%%%%%%%%%%%%%%%%%%%%%%%%%%%%%%%%%%%%%%%%%%%%%%%%%%%%%%
The mass-energy  and angular momenta or these solutions
is computed by using the procedure proposed by  Balasubramanian
and Kraus \cite{Balasubramanian:1999re},
which furnishes a means for calculating the
gravitational action and conserved quantities
without reliance on any reference spacetime.
This technique was inspired by AdS/CFT correspondence and consists
of adding suitable counterterms $I_{ct}$
to the action of the theory in order to ensure the finiteness of the boundary
stress tensor \cite{Brown:1993br}.
As found in \cite{Balasubramanian:1999re},
the following counterterms are sufficient to cancel
divergences in five dimensions,
for AdS$_5$ vacuum black hole solutions\footnote{These counterterms
regularize also the mass-energy and  action of rotating Einstein-Maxwell-AdS solutions
in  \cite{Kunz:2007jq}.}
(here $\rm{R}$ is the Ricci scalar for the boundary metric $h$)
\begin{eqnarray}
\label{ct} I_{\rm ct}=-\frac{1}{8 \pi G} \int_{\partial {\cal
M}}d^{4}x\sqrt{-h}\Biggl[ \frac{3}{ \ell}+\frac{ \ell}{4}\rm{R}
\Bigg]\ .
\end{eqnarray}
Using these counterterms one can
construct a divergence-free stress tensor from the total action
$I_{tot}{=}I{+}I_{\rm ct}$ by defining
\begin{eqnarray}
\label{s1}
{\rm T}_{AB}&=& \frac{2}{\sqrt{-h}} \frac{\delta I_{tot}}{ \delta h^{AB}}
=\frac{1}{8\pi G}(K_{AB}-Kh_{AB}-\frac{3}{\ell}h_{AB}+\frac{\ell}{2} E_{AB}),
\end{eqnarray}
where $E_{AB}$ is the Einstein tensor of the intrinsic metric $h_{AB}$.

The presence of the additional matter fields in the bulk action
brings the potential danger of having divergent contributions coming
from both the gravitational and matter action
\cite{Taylor-Robinson:2000xw}. Various examples of
AAdS solutions whose action and mass {\it cannot} be regularized by
employing only the counterterm (\ref{ct}) have been presented in the
literature.
This is also the case of the AdS$_5$ non-Abelian solutions, where the backreaction
of the gauge fields causes certain metric components
to fall off slower than usual.
As a result, the action and the mass-energy
 present generically a logarithmic divergence,
unless one considers
corrections to the YM Lagrangean consisting of higher order terms of the
Yang--Mills hierarchy \cite{Radu:2005mj}.

However, in such cases, it is still  possible to obtain a finite
mass and action by allowing $I_{ct}$ to depend not only on the
boundary metric $h_{AB}$, but also on the matter fields.
The  matter counterterm expression which is added
to $I_{tot}$ for  AdS$_5$ non-Abelian solutions  is \cite{BRT}
(with $A,B$ boundary indices)
\begin{eqnarray}
\label{Ict-mat}
I_{ct}^{(m) }=
-\log{\left(\frac{r}{\ell}\right)}
 \int_{\partial M}d^{4}x\sqrt{-h }\frac{\ell}{2e^2}
 ~{\rm tr}\{ F_{AB}F^{AB}\}~,
\end{eqnarray}
which yields a supplementary contribution to (\ref{s1})
\begin{eqnarray}
\label{TAB-mat}
{\rm T}_{AB}^{(m)}=-\log{\left(\frac{r}{\ell}\right)} \frac{2\ell}{e^2}
~{\rm tr}\{F_{AC}F_{BC}h^{CD}-\frac{1}{4}h_{AB}~F_{CD}F^{CD}\}.
\end{eqnarray}
Provided the boundary geometry has an isometry generated by a
Killing vector $\xi$, a conserved charge
\begin{eqnarray}
{\frak Q}_{\xi }=\oint_{\Sigma }d^{3}S^{i}~\xi^{j}{\rm T}_{ij}
\label{charge}
\end{eqnarray}
can be associated with a closed surface $\Sigma $ \cite{Balasubramanian:1999re}.
If $%
\xi =\partial /\partial t$ then ${\frak Q}$ is the conserved
mass/energy $E$; there are also two angular momenta associated
with the Killing vectors $\partial/\partial \varphi$ and $\partial/\partial \psi$.

As a result, we find the
following expressions for mass-energy  and
angular momentum of the solutions in this paper\footnote{Note that these quantities are
evaluated in a frame which is nonrotating
at infinity.}:
\begin{eqnarray}
\label{MT-rot}
&&E= -\frac{V_3}{8\pi G}
 \left(\frac{3f_2}{2}+\frac{4s_4}{\ell^2}\right)+E_c,~~
J_\varphi=J_\psi=J=-\frac{\hat{J}V_3}{16\pi G} ~,
\end{eqnarray}
where $E_c=3\pi\ell^2/32 G$
is a constant terms
interpreted as the mass-energy of the AdS$_5$ background \cite{Balasubramanian:1999re}
and $V_3=2\pi^2$ is the area of the three sphere.
One can prove that the term (\ref{Ict-mat})
regularizes also the tree level action of the solutions\footnote{In the absence of closed form
solutions, there is no obvious way to perform
a meaningful Wick rotation and obtain a real Euclidean solution
for a rotating non-Abelian black hole.
However, one can use a quasi-Euclidean approach as described
in \cite{quasi}. }.

%%%%%%%%%%%%%%%%%%%%%%%%%%%%%%%%%%%%%%%%%%%%%%%%%%%%%%%%%%%%%%%%
\subsubsection{Other relations}
%%%%%%%%%%%%%%%%%%%%%%%%%%%%%%%%%%%%%%%%%%%%%%%%%%%%%%%%%%%%%%%%
These solutions have also an electric charge
\begin{eqnarray}
\label{Q_e} Q_e= -\frac{1}{V_3}\lim_{r \to \infty}\int dS_k { {\rm
tr}} \{F^{kt}\frac{\tau_3}{2}\}=q.
\end{eqnarray}
By using the fact that the integral of the angular momentum density
 can be written as a difference of two
boundary integrals \cite{VanderBij:2001nm}, one writes
\begin{eqnarray}
\label{rel1}
 \int d^4x~T_\varphi^t \sqrt{-g} =
\oint_{\infty}2{\rm tr}\{A_\varphi F^{\mu t} \}
dS_{\mu}-\oint_{r=r_h}2{\rm tr}\{A_\varphi F^{\mu t} \} dS_{\mu}.
\end{eqnarray}
(a similar relation holds for $T_\psi^t$).
Making use of the Einstein equations, one finds
the following relation
\begin{eqnarray}
\label{rel2}
J-2Q_e(W_0-1)=\frac{g_hr_hH_h}{\sigma_h}
\bigg( V'(r_h)+2\Omega_H H'(r_h) \bigg)
+\frac{g_hr_h^3\Omega'(r_h)}{8\sigma_h}~,
\end{eqnarray}
relating  global charges to event horizon quantities.
It is also of interest to evaluate
 the integral of  ${\rm tr}\{ F_{\mu t} F^{\mu t} \}$.
This measures the contribution of the non-Abelian electric field to
the mass/energy of the system. Similar to the four dimensional case,
by using the YM equations this integral can be expressed as
\begin{eqnarray}
\label{electric-mass2} -E_e =\int {\rm tr}\{ F_{\mu t} F^{\mu t}
\}\sqrt{-g}d^4x=
 \oint_{\infty} {\rm tr} \{A_t F^{\mu t} \} dS_{\mu}-\oint_{eh} {\rm tr} \{A_t F^{\mu t} \} dS_{\mu}.
\end{eqnarray}
Thus, for globally  regular configurations,
a vanishing magnitude of the electric potentials at infinity
implies a purely magnetic solution.
In contrast, one finds rotating black
hole solutions with $A_t(\infty)=0$ which are supported by the
event horizon contribution.
Since the asymptotic expansion (\ref{a1})
 holds for both globally regular and black hole solutions,
 we conclude that there are no rotating EYM-SU(2) solitons
in AdS$_5$
(the condition $V(\infty)=0$ follows from the physical
requirement
that the spacetime approach the AdS background at infinity).
However, rotating soliton solutions are likely to exist for a larger gauge group.

%%%%%%%%%%%%%%%%%%%%%%%%%%%%%%%%%%%%
\begin{figure}[h!]
\parbox{\textwidth}
{\centerline{
\mbox{
\epsfysize=15.0cm
\includegraphics[width=82mm,angle=0,keepaspectratio]{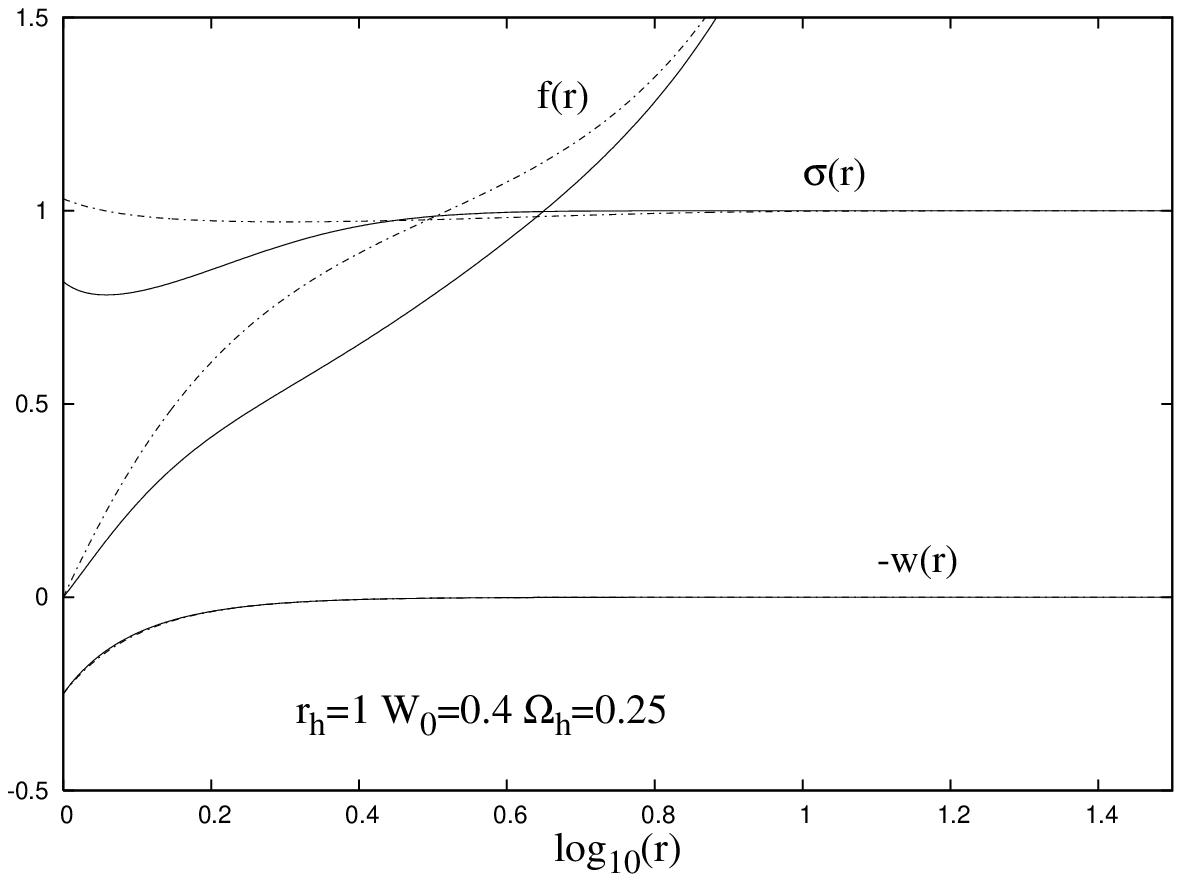}
\includegraphics[width=82mm,angle=0,keepaspectratio]{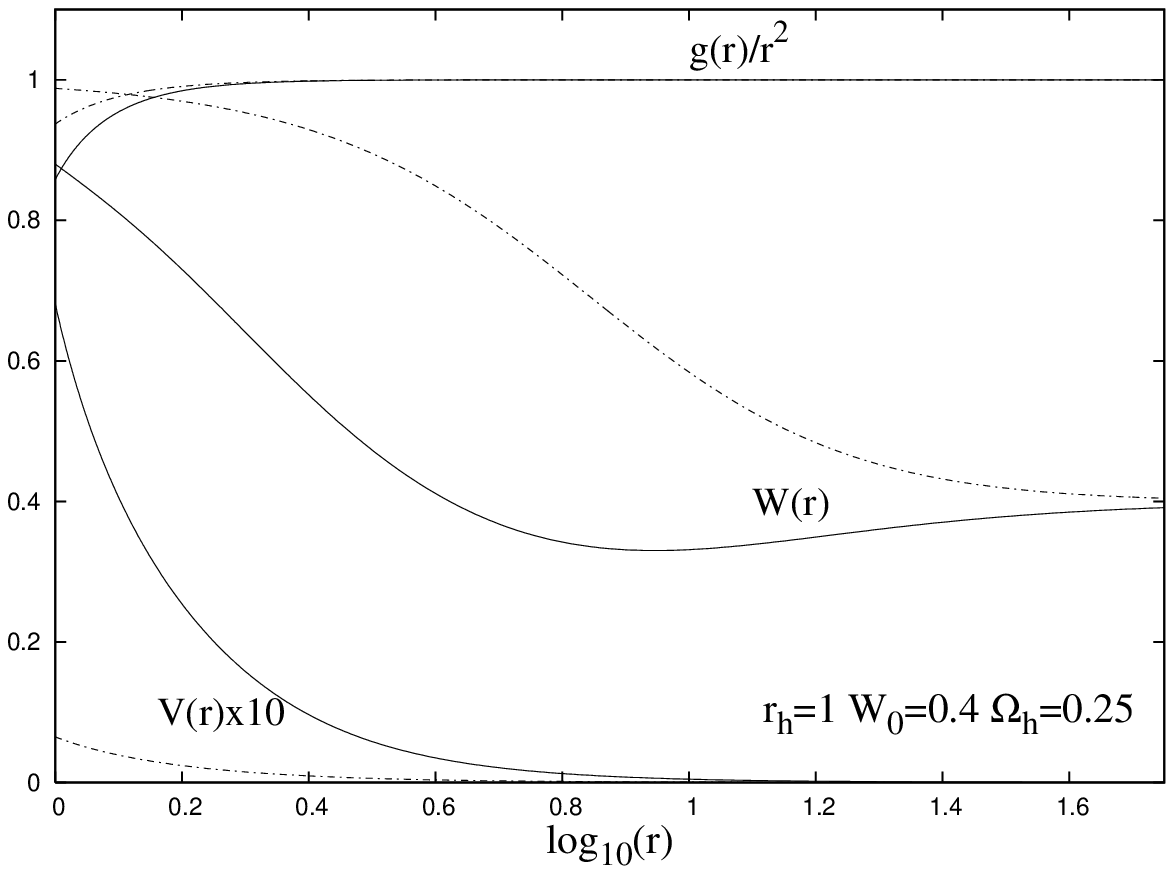}
}}}
\caption{{\small The profiles of the metric functions $f(r)$, $g(r)/r^2,~w(r),~\sigma(r) $ and
 the non-Abelian gauge potentials
 $W(r),~V(r)$ are shown for two typical
charged rotating black hole solutions with the same values of event horizon radius $r_h$,
event horizon angular velocity $\Omega_H$  and
magnitude of the magnetic potential at infinity $W_0$. }}
\end{figure}
%%%%%%%%%%%%%%%%%%%%%%%%%%%%%%%%%%%%%%%

The Killing vector  $\chi=\partial/\partial_t+
 \Omega_\varphi \partial/\partial \varphi + \Omega_\psi \partial/\partial \psi $ is
orthogonal to and null on the horizon. For the solutions
within the ansatz (\ref{metric}), the
event horizon's angular velocities are
all equal, $\Omega_\varphi=\Omega_\psi=w(r)|_{r=r_h}$.
The Hawking temperature as  found by computing
the surface gravity is
\begin{eqnarray}
\label{Temp-rot}
  T_H=\frac{\sqrt{b'(r_h)f'(r_h)}}{4\pi}.
\end{eqnarray}
Another quantity of interest is
the area $A_H$ of the rotating black hole horizon
\begin{eqnarray}
\label{A2}
A_H= r_h  g_h V_3 .
\end{eqnarray}
As usual, one identifies the entropy of black hole solutions with
one quarter of the even horizon area, $S=A_H/4G$.

To have a measure of the deformation of the horizon, we introduce a
deformation parameter defined as the ratio of the equatorial circumference
$L_e$ and
the polar one $L_p$, which for these solutions we are considering takes the
form
\begin{eqnarray}
\label{exc}
\frac{ L_e}{L_p} =  \frac{r_h} {\sqrt{g(r_h)}} \ .
\end{eqnarray}
These rotating solutions present also an ergoregion inside of which
the observers cannot remain stationary, and will move in the
direction of rotation. The ergoregion is the region bounded by the
event horizon, located at $r=r_h$ and the stationary limit surface,
or the ergosurface, $r=r_e$.
 The Killing
vector $\partial/\partial t$ becomes null on the  ergosurface ,
$i.e.$ $g_{tt}(r_e)= -b(r_e)+r^2 w^2(r_e)=0$. The  ergosurface does
not intersect the horizon.

%%%%%%%%%%%%%%%%%%%%%%%%%%%%%%%%%%%%%%%%%%%%%%%%%%%%%%
\section{The properties of solutions}
%%%%%%%%%%%%%%%%%%%%%%%%%%%%%%%%%%%%%%%%%%%%%%%%%%%%%%

Although we have considered other values as well, the numerical results reported
in  this section corresponds to $\ell=10$, which is also the value taken in the study
 \cite{Okuyama:2002mh} of the static solutions.
%
%%%%%%%%%%%%%%%%%%%%%%%%%%%%%%%%%%%%%%%%%%%%%%%%%%%%%%
\begin{figure}[h!]
\parbox{\textwidth}
{\centerline{
\mbox{
\epsfysize=15.0cm
\includegraphics[width=112mm,angle=0,keepaspectratio]{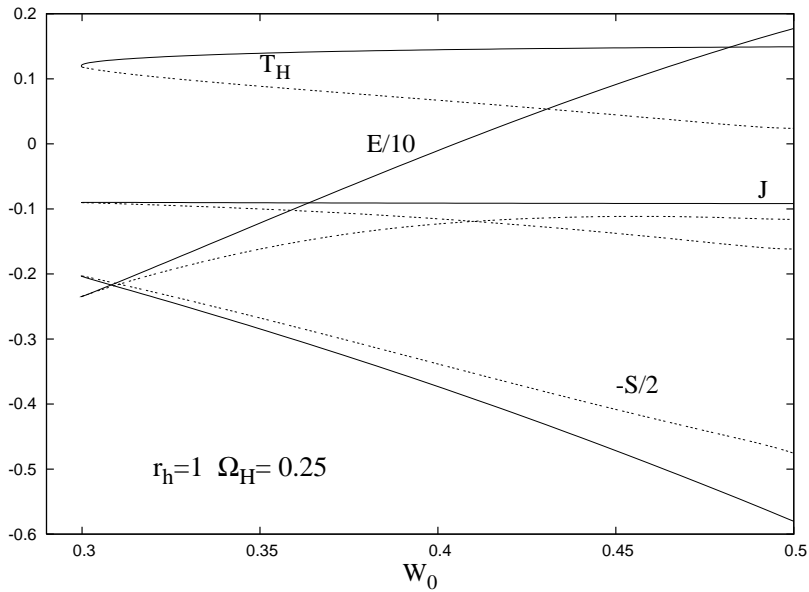}
}}}
\caption{{\small
Some relevant parameters are plotted
as a function of $W_0$ (with $W_0$ near  the critical value $W^{(cr)}_0$)
for rotating black hole solutions with $r_h=1$, $\Omega_H=0.25$.}}
\end{figure}
%%%%%%%%%%%%%%%%%%%%%%%%%%%%%%%%%%%%%%%%%%%%%%%%%%%%%%%%%
Dimensionless quantities are obtained by using the
rescaling $r \to r \sqrt{4\pi G}/e  $,
and $\Lambda \to \Lambda e^2/(4\pi G)$.
To integrate the equations, we used the differential
equation solver COLSYS which involves a Newton-Raphson method
\cite{COLSYS}.

We start the description here by recalling the situation in the static limit.
Spherically symmetric non-Abelian black holes
exist for any value of the event horizon radius, a
globally regular configuration being approached
as $r_h\to 0$.
The parameter $W_0$ in the boundary conditions at infinity
is not fixed; however, one finds the existence of a minimal value
of $W_0$, which depends on $r_h$.
The mass of spherically symmetric black holes
as defined in (\ref{MT-rot}) may take negative values as well,
for a range of ($W_0,~r_h$).
One notice also the possible existence of several configurations
for the same set $(W_0,r_h)$.

The rotating solutions we have found preserve  this general picture.
As expected, we could find rotating solutions starting with any
static black hole. Rotating solutions are obtained by increasing the
value of $\Omega_H$ or $\hat J$ \footnote{In the numerical
procedure, we have fixed the values of $W_0$, $\Omega_H$ (or $\hat
J$) together with $V(r_h)=-2\Omega_H H_h$, $f(r_h)=0$, and a  set of
three more complicated conditions at the horizon involving both the
functions and their derivatives.}.
For all the solutions
we studied, the metric functions $f(r)$, $g(r)$ , $\sigma(r)$
and $w(r)$ interpolate
 monotonically between the corresponding values at $r=r_h$ and the
asymptotic values at infinity, without developing  any pronounced
local extrema. (The magnetic gauge potentials present, however, a
more complicated behaviour.) As a typical example,  we present in
Figure $1$  the profile of two  solutions with the same values of
$r_h,~W_0=0.3$ and $\Omega_H$. These configurations are clearly
distict and have different global charges.

The basic geometrical features of these rotating solutions are rather
similar to the vacuum or U(1) case
($e.g.$ the presence of an ergosphere and the fact
that the horizon is deformed\footnote{However, different from the
U(1) case \cite{Kunz:2007jq},
the rotating non-Abelian solutions we have studied have $L_e/L_p>1$ only.}).
%
%%%%%%%%%%%%%%%%%%%%%%%%%%%%%%%%%%%%%%%%%%%%%%%%%%%%%%%%%
\begin{figure}[h!]
\parbox{\textwidth}
{\centerline{
\mbox{
\epsfysize=15.0cm
\includegraphics[width=82mm,angle=0,keepaspectratio]{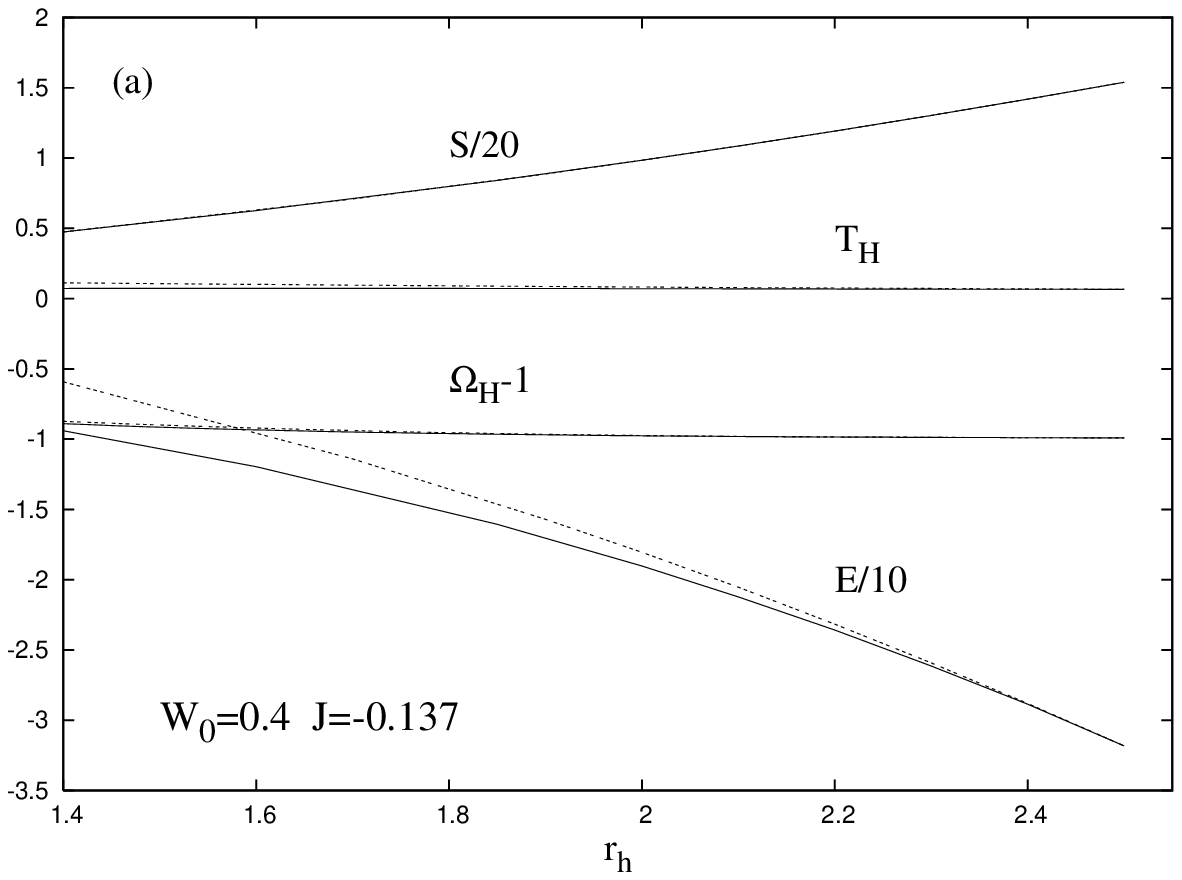}
\includegraphics[width=82mm,angle=0,keepaspectratio]{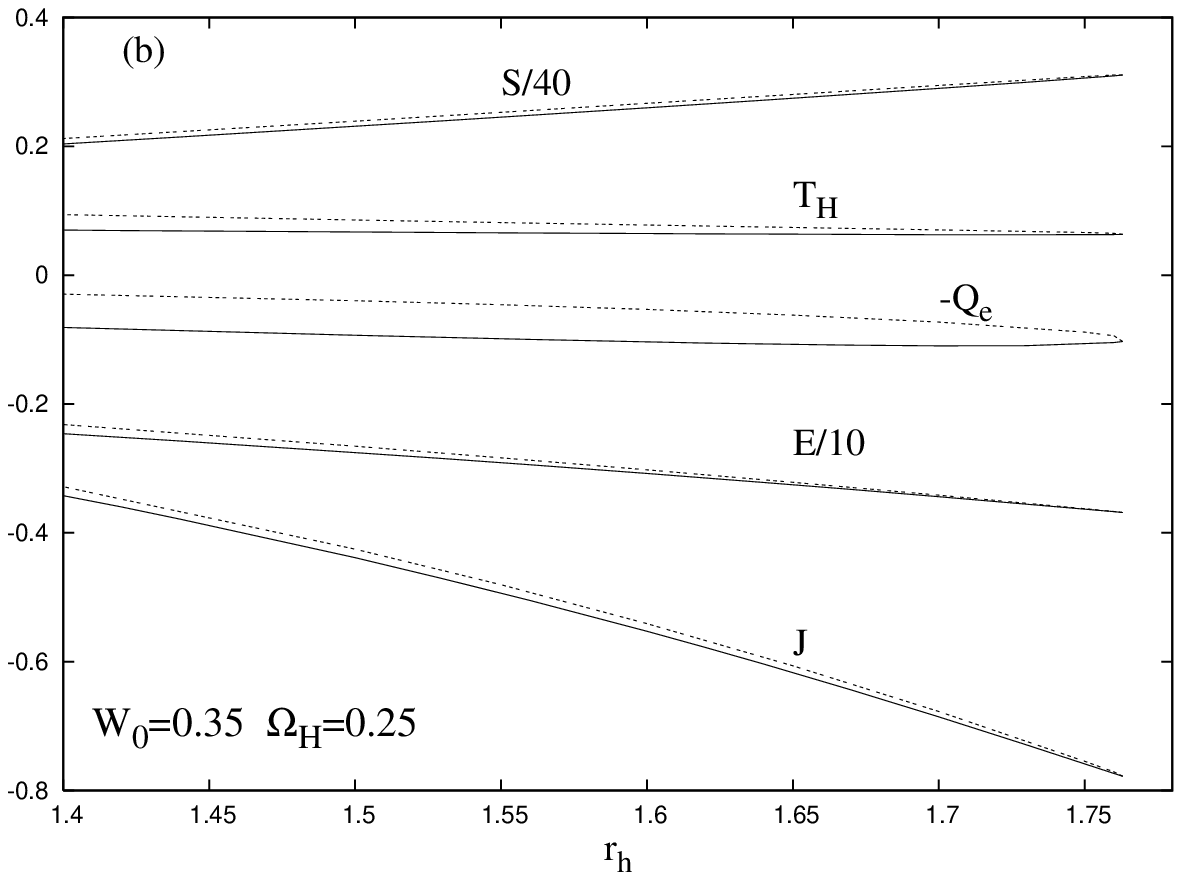}
}}}
\caption{{\small Some relevant parameters
are plotten as a function of event horizon radius for
rotating black holes with a fixed value of
magnetic
potentials at infinity and the same angular momenta (a) or event horizon velocity (b). }}
\end{figure}
%%%%%%%%%%%%%%%%%%%%%%%%%%%%%%%%%%%%%%%%%%%%%%%%%%%%%%%%%
%
In the numerics we have paid special attention to the solutions' dependence
on   the magnitude of the magnetic potentials at infinity $W_0$,
which is a purely non-Abelian feature.
Even in this case, the configurations present a very rich structure, which makes
their complete classification in the
space of physical parameters
a considerable task which is not aimed in this paper.
Instead, we analyzed
in details a few particular classes of solutions which, hopefully,
reflect all the properties of the general pattern.

A feature of the rotating solutions we have studied
so far is the existence of two different solutions
for the same values of ($r_h,\Omega_H,W_0$).
These solutions have different global charges and distinct temperatures.
No upper limit on $W_0$ seems to exist (although the numerics
become very difficult for large $W_0$).
 When fixing the
event horizon radius and the rotation parameter $\Omega_H$ (or $\hat
J$), we have noticed, similar to the static case, the existence of a
minimal value of $W_0$, say $W_0^{(cr)}(r_h,\Omega_H)$. At that
point, a secondary branch of solutions emerges, which extends to
larger values of $W_0$. This behaviour is illustrated in Figure 2.
The occurrence of a minimal value of $W_0>0$ makes it unlikely that
the non-Abelian black holes constructed here are bifurcations of the
Abelian solutions which correspond to setting $W=0$ in the
equations.

We have also studied the  dependence of solutions properties on the
value of the event horizon radius for fixed $w_0$ and $J$. The
numerical results strongly support the existence of two branches of
rotating black hole solutions which join at a maximal value of $r_h$
(see Figure 3a). This result further suggests that no solution exist
for higher values of the event horizon $r_h$.
 As expected, the same pattern is found when taking instead a
 constant value of the event horizon velocity instead of $J$,
 see Figure 3b.

When a rotating black hole solution is considered for
$r_h \to 0$ with the other parameters fixed, we observe
that  $V(r_h)$ converge to zero (in fact the electric
potential $V$ tends to zero uniformly in this
limit) as well as $J$. In the same time the value
$\vert w'(r_h) \vert$ goes to infinity, so that
the rotation function $w(r)$ becomes more and more
peaked at $r=r_h$.
%
%
%%%%%%%%%%%%%%%%%%%%%%%%%%%%%%%%%%%%
\begin{figure}[h!]
\parbox{\textwidth}
{\centerline{
\mbox{
\epsfysize=15.0cm
\includegraphics[width=82mm,angle=0,keepaspectratio]{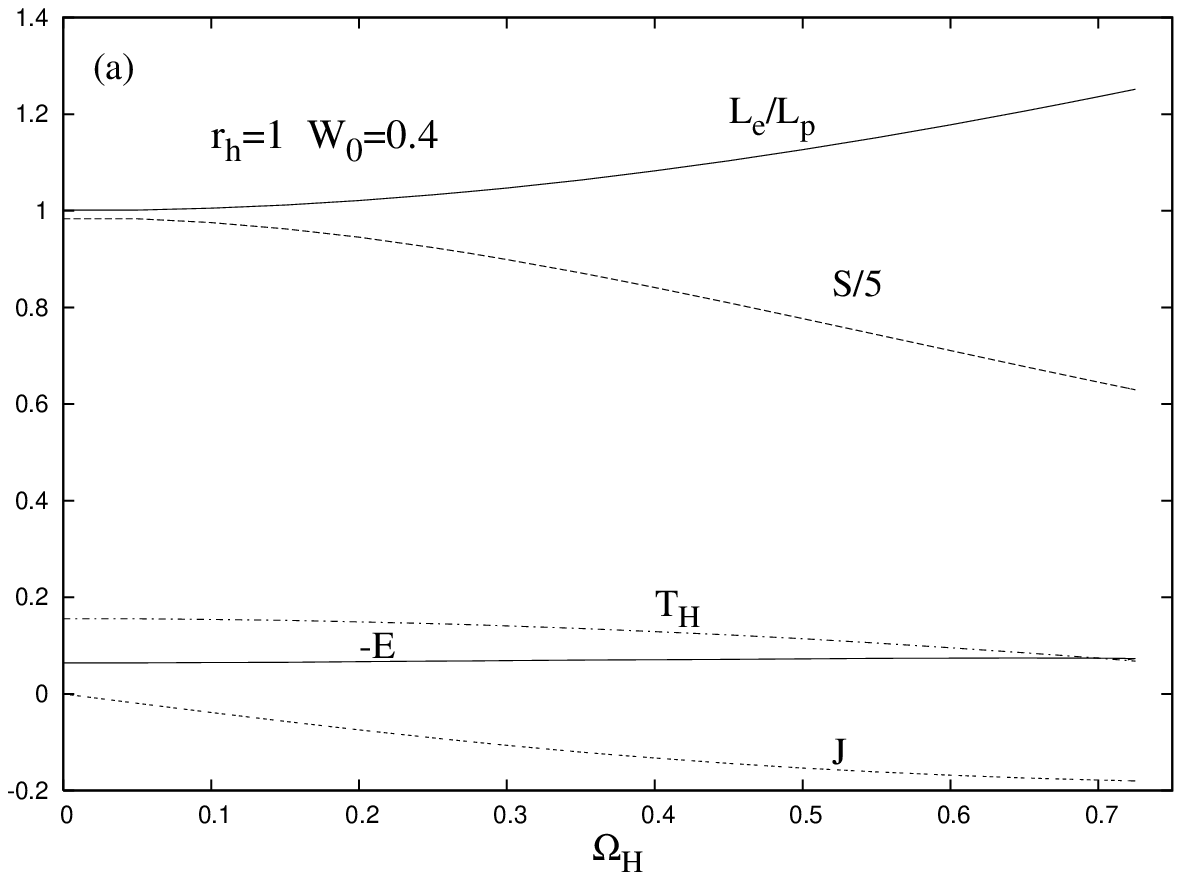}
\includegraphics[width=82mm,angle=0,keepaspectratio]{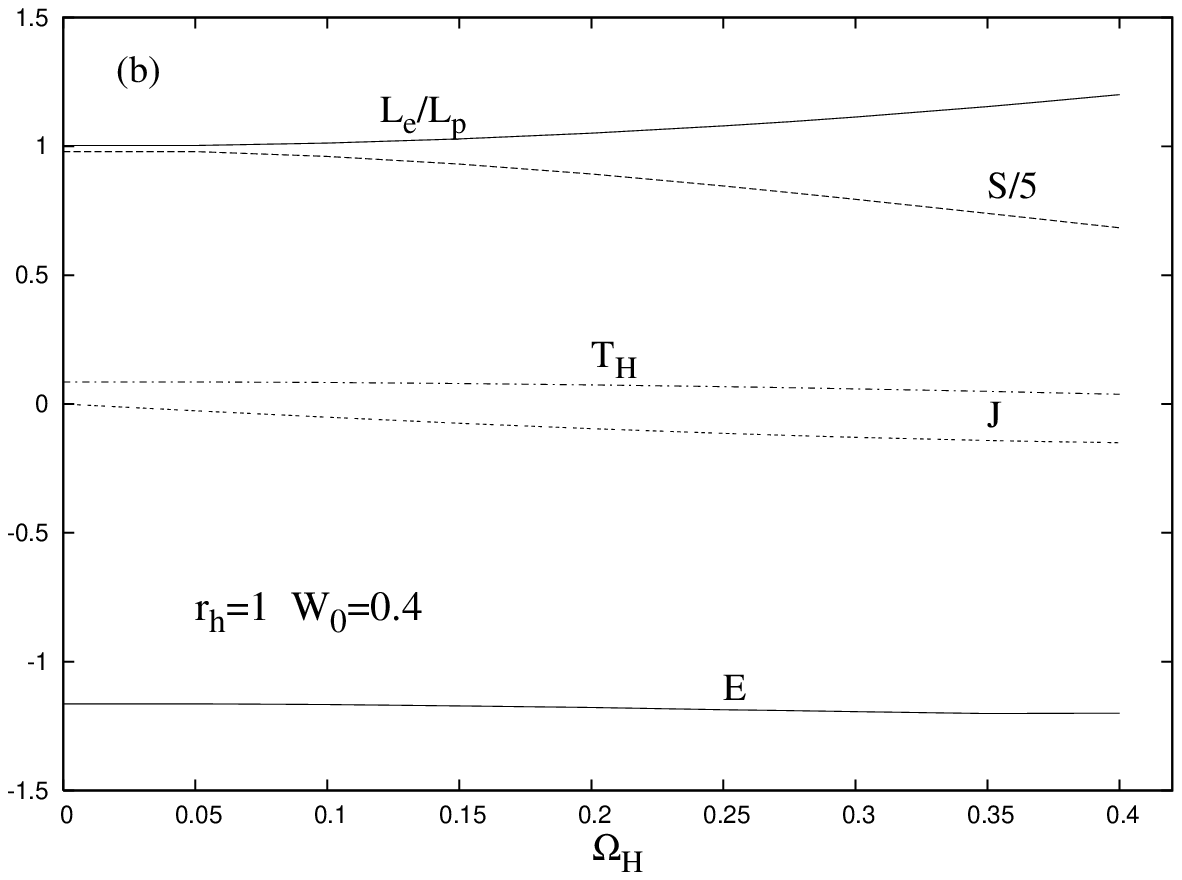}
}}}
\caption{{\small The evolution of some relevant parameters
is plotten as a function of event horizon velocity for two branches
rotating black holes with the same event horizon radius $r_h$
and the same values of  the
 magnetic potential  at infinity.}}
\end{figure}
%%%%%%%%%%%%%%%%%%%%%%%%%%%%%%%%%%%%%%%
As a result, no rotating soliton solution is found.

Finally, we examine the properties of the solutions on the value
$\Omega_H$ of the event horizon velocity. The numerical results
obtained suggest that $W_0^{(cr)}$ depends weakly on $\Omega_H$.
Similar to the vacuum of EM cases, we noticed here again the
existence of two branches of solutions. However, no definite
conclusion is unfortunately available due to severe numerical
difficulties that we met when increasing the value of $\Omega_H$.
The evolution of different parameter characterizing this family of
solutions is reported on Figure 4 for $r_h=1, W(r_h)=0.4$; as far as
we could see there is no signal that the two branches will meet at a
maximal value of $\Omega_H$, as happens for vacuum or EM rotating
solutions (for $r_h=1,~W_0 = 0.4$ we could integrate the first
branch up to $\Omega_H \simeq 0.75$ and the second branch up to
$\Omega_H \simeq 0.45$, the numerical results becoming unreliable
for larger values of $\Omega_H$). This  suggests that the two
branches remain may open and exist for large values of the angular
momentum. Integration of the equations with a different technique
and/or a different metric parametrization may clarify this issue.

%%%%%%%%%%%%%%%%%%%%%%%%%%%%%%%%%%%%%%%%%
\section{Further remarks}
%%%%%%%%%%%%%%%%%%%%%%%%%%%%%%%%%%%%%%%%%
The main purpose of this paper was to present arguments for the
existence of a general class of five dimensional AdS charged
rotating solutions in EYM theory, in which the two angular momenta
are equal. These solutions depend on four nontrivial parameters,
namely the the mass, the angular momenta, the electric charge and
the essential value of a magnetic potential at infinity.

This class of solutions may provide a fertile ground for further
study of charged rotating configurations in gauged supergravity
models and one expects some of their properties to be generic. Our
preliminary results indicate the presence of similar solutions in
the ${\cal{N}}=4^+$ version of the Romans' gauged supergravity
model, with a dilaton potential presenting a stationary point
\cite{Romans:1985ps}. Rotating EYM topological black holes with an
horizon of negative curvature are also likely to exist for
$\Lambda<0$. In addition, it would be interesting to generalize
these solutions to higher dimensions, thus extending the study of
charged U(1) black holes in \cite{Kunz:2007jq} to EYM-$\Lambda$
theory.

The study of the solutions discussed in this paper in an AdS/CFT
context is an interesting open question. A generic property of the
non-Abelian fields in AAdS backgrounds is that they do not approach
asymptotically a pure gauge configuration. The boundary form of the
non-Abelian potential (\ref{ansatz-gauge}) is
\begin{eqnarray}
\label{A0}
A_{(0)}= W_0\tau_1d \theta +\left(- W_0\sin 2 \theta\frac{1}{2}\tau_2+
 ( -W_0\cos 2 \theta +W_0-1)\frac{1}{2}\tau_3\right) d \varphi+
\\
\nonumber
\left(W_0\sin 2 \theta\frac{1}{2}\tau_2+
 (W_0\cos 2 \theta +W_0-1)\frac{1}{2}\tau_3\right) d \psi~,
\end{eqnarray}
with a nonzero boundary field strength tensor  $F_{ (0)}^{\mu \nu}$
(note that $A^{(0)}$ can be gauged away in the Abelian limit
$W_0=0$). On the CFT side, these fields corresponds to external
source currents coupled to various operators.

The metric on which the boundary CFT is defined is found by as
$\gamma_{ab}=\lim_{r \rightarrow \infty} \frac{\ell^2}{r^2}h_{ab}$,
  and corresponds to
a static Einstein universe in four dimensions,
\begin{eqnarray}
\label{b-metric}
\gamma_{ab}dx^a dx^b=-dt^2+\ell^2d\Omega^2_3.
\end{eqnarray}
One can use the AdS/CFT ``dictionary'' to predict
qualitative features of a quantum field theory in this  background.
For example,  the expectation value of the dual CFT stress-tensor
can be calculated using the  relation \cite{Myers:1999qn}
%\begin{eqnarray}
%\label{r1}
$\sqrt{-\gamma}\gamma^{ab}<\tau_{bc}>=
\lim_{r \rightarrow \infty} \sqrt{-h} h^{ab}{\rm T}_{bc}.$
%\end{eqnarray}
For these solutions we find the
following non-vanishing components of the dual CFT
stress-energy tensor
(with $x^1=\theta,~x^2=\varphi,~ x^3=\psi,~x^4=t$)
\begin{eqnarray}
  <\tau^{a}_b> = \frac{N^2}{4 \pi^2 \ell^4}\bigg[
\frac{1}{2 }
\big(
\frac{1}{4}
-\frac{f_2}{\ell^2}-\frac{4s_4}{\ell^4}
\big)\left( \begin{array}{cccc}
1&0&0&0
\\
0&1&0&0
\\
0&0&1&0
\\
0&0&0&-3
\end{array}
\right)
+
\frac{2s_4}{ \ell^4} \left( \begin{array}{cccc}
0&0&0&0
\\
0&\sin^2 \theta &\sin^2 \theta  &0
\\
0&\cos^2 \theta &\cos^2 \theta  &0
\\
0&0&0&-1
\end{array}
\right)
\\
\nonumber
 +
 2\hat J \left( \begin{array}{cccc}
0&0&0&0
\\
0&0&0&\sin^2 \theta
\\
0&0&0&\cos^2 \theta
\\
0&-\frac{1}{\ell^2}&-\frac{1}{\ell^2}&0
\end{array}
\right)
\bigg]
-\frac{8}{e^2 }\frac{ (W_0-1)^2 W_0^2}{\ell^3}
\left( \begin{array}{cccc}
1&0&0&0
\\
0&1&0&0
\\
0&0&1&0
\\
0&0&0&0
\end{array}
\right)
,
\end{eqnarray}
where we have replaced $8\pi G=4\pi^2\ell^3/N^2$
\cite{Maldacena:1997re}, with $N$ the rank of the gauge group of the
dual $\mathcal{N}=4,~d=4$ theory. The first three terms in this
relation appear also for other known rotating solutions with equal
magnitude angular momenta. The last term, however, is due to the
existence of a non-Abelian matter content in the bulk and implies a
nonvanishing trace of the $d=4$ CFT stress tensor, $<\tau^{a}_a>=-24
(W_0-1)^2 W_0^2/(\ell^3 e^2)$. From (\ref{A0}), this can be written
as $<\tau^{a}_a>=-\frac{\ell }{4e^2}F_{(0)}^2$, in agreement with
the general results \cite{BRT}.

Further progress in this direction may require to embed these
solutions in a supergravity model.

\medskip
\medskip
%%%%%%%%%%%%%%%%%%%%%%%%%%%%%%%%%%%%%%%%%%%%%%%%%%%%%%%%%%%%%%%%%%%%%%%
\noindent
{\bf\large Acknowledgements}\\
YB is grateful to the
Belgian FNRS for financial support.
The work of ER and DHT was carried out
in the framework of Science Foundation--Ireland (SFI)
 Research Frontiers Programme (RFP) project
RFP07/FPHY330.

\begin{small}

\end{small}

\end{document}